# AI-Driven Feedback Systems, Digital Labour, and Silent Quitting: Transforming African Workplaces

**Abayomi O. Agbeyangi & Jose M. Lukose**
Department of Business and Applications Development, Walter Sisulu University,
East London, South Africa
*Corresponding: aagbeyangi@wsu.ac.za*

**Abstract**

The current trend of digitalisation has revolutionised the organisation of work and the way it is measured and performed across the globe, with AI becoming more common for managing labour and performance, as well as employee communication. In African organisations, where there is increasing adoption of remote work, hybrid models of work, digital collaboration, and data-based HR management, the notion of silent quitting has become more relevant, defined as worker disengagement when employees are still doing their job but do not put any effort into achieving good performance and exhibiting any emotion. This paper investigates how AI-driven feedback mechanisms, including sentiment analysis systems, pulse surveys, chatbots, engagement dashboards, and predictive analytics, are changing African workplaces through offering continuous listening, instant performance information and proactive engagement with employees. The study also explores how AI can assist organisations in identifying early disengagement and enable intervention and better employee communication in both private and public sector organisations in Africa. At the same time, we address the challenges of socioeconomic development and governance posed by AI implementation in developing countries, including digital inequality, infrastructure shortcomings, privacy concerns, algorithmic bias, and the risk of workplace surveillance. By situating silent quitting within wider debates on digital labour and automation, the paper contributes an African-centred perspective to discussions on the future of work and offers practical recommendations for HR professionals, managers, policymakers, and technology developers seeking responsible, context-sensitive approaches to workplace transformation across the continent.

## 1. Introduction

Silent quitting (or quiet quitting (QQ)) refers to the state whereby employees disengage emotionally from their tasks, yet continue meeting minimum requirements at work (Dutta *et al.*, 2024; Gabelaia and Bagociunaite, 2024; Gone *et al.*, 2025; Gün *et al.*, 2025). Silent quitting has become increasingly popular as a global phenomenon in the modern world, as it has a wide array of repercussions on efficiency, mood, and organisational unity. Indeed, as employees detach themselves from their work psychologically and stop showing discretionary efforts at work, it becomes challenging for employers to deal with the situation because all the negative aspects become apparent only when performance and trust have already been damaged. Although there are a number of studies that explore the phenomena of burnout and poor leadership in developed countries, there is limited research on how silent quitting occurs in the African context, where economic insecurity, uneven digital transformation, and variable labour protections shape distinctive engagement conditions (Georgiadou *et al.*, 2025)

On the other hand, there is an ongoing structural change happening within the workplace. All over Africa, both public and private firms have started using digital collaboration systems, hybrid working schemes, cloud-based performance systems, and other HR technology driven by data analytics (Amoako et al., 2023). This reflects the larger process of digital labour, where the labour becomes more organised, monitored, evaluated, and regulated via digital technologies. In this way, the AI-based feedback systems for HR are not just technological means for improving efficiency, but one of the elements of the wider restructuring of labour regulation, where the workers are supervised, assessed, assisted, and even heard (Chilunjika et al., 2022; Mamuli et al., 2025). As the digital systems become mediators for communication and performance, they change the relations of power, affect the degree of autonomy of the workers, and give rise to new tensions between assistance and supervision.

Against this backdrop of evolving trends, the concept of silent quitting can be viewed not only in terms of the psychological phenomenon resulting from burnout or job dissatisfaction, but also as a labour phenomenon arising amid digitally transformed working conditions. Post-pandemic adoption of remote and hybrid work arrangements has created confusion between work and private life, created the expectation of availability around the clock, and thus made disengagement of the workforce hard to spot and interpret correctly (Cavicchioli et al., 2025; Yao et al., 2024). This problem becomes even more pressing within many African organisations due to infrastructural problems, poor accessibility of digital solutions, HR digitalisation, and the culture which discourages any upward communication. Employees become disengaged silently while their managers do not even know about the true reasons behind this (Govender & Bussin, 2020; Rivera et al., 2021).

Emerging technologies such as real-time sentiment analysis (Lee and Song, 2024), engagement dashboards (Ramu et al., 2026), AI-driven pulse surveys, and automated feedback loops present innovative means of gauging employee concerns and responding to them before they become firmly established. Through machine learning and natural language processing, they allow for pattern recognition in employees' communications, early detection of morale decline, and enable more precise interventions (Madhumita et al., 2024; Sampath et al., 2024; Wendakoon, 2024). For African organisations, characterised by limited resources and hierarchical structures that make it difficult to conduct feedback efficiently, these AI-based tools could prove valuable in the context of listening and acting on a timely basis. Yet their implementation poses several ethical issues related to privacy, equity, cultural sensitivity, trust, and digital divide (Ara and Ahmad, 2025; Madanchian et al., 2023).

This paper argues that AI-driven feedback systems should be examined through the combined lenses of employee engagement, digital labour, and workplace transformation. It explores how such technologies can help identify early signals of silent quitting, while also considering how they may deepen managerial control, reinforce surveillance, or widen inequalities if poorly governed. It, therefore, positions AI as both an opportunity and a governance challenge in African workplaces. By focusing on the intersection of automation, employee disengagement, and organisational change, the paper contributes an African-centred perspective to broader debates on the socioeconomic impacts of AI in the Global South.

The structure of the paper is as follows: Section 2 looks at the evolution of silent quitting in the digital era and its connection to changes in digital labour. Section 3 examines the use of AI technologies in the context of employee feedback and engagement, and shows how these technologies are transforming labour management. Section 4 offers some African workplace examples to showcase how new forms of digital transformation and responses of employees can look like. Section 5 addresses the ethical, cultural, technical and socio-economic barriers to introducing AI-powered feedback solutions in African firms. Section 6 gives some suggestions on how to implement AI ethically and culturally sensitive. Section 7 concludes the paper.

## 2. Understanding Silent Quitting in the Digital Age

### 2.1 Historical Roots and Evolution

Silent quitting or quiet quitting (*both terms used interchangeably in this paper*) differs from actual resigning in that it entails emotional detachment, wherein employees perform their basic tasks without any discretionary actions. Silent quitting is not a recent occurrence in the history of labour and workplace behaviours. Rather, historically speaking, it has been an advanced stage of a preexisting type of industrial action, which is termed "work to rule." Bloch and Moorman (1993); Hamouche et al. (2023) assert that "work-to-rule" is a type of tactic historically used by trade unions to oppose or complain about the working conditions through workers following every single procedure and rule associated with their jobs to protest against the employers. By performing their jobs in accordance with what is clearly stated in their job descriptions, workers were able to bring disruptions to organisations as a form of collective resistance. The "work to rule" effect, according to Hamouche et al. (2023), is a concept of industrial relations wherein employees do the minimum that they are required by their job descriptions as part of a collective action for demanding higher pay and improved working conditions. Table 1 presents some identified differences between silent quitting and work-to-rule.

What distinguishes the current phenomenon of quiet quitting is the stealth and personalisation of this kind of behaviour. While work-to-rule tactics were usually a direct form of protest employed by workers during labour disagreements, quiet quitting is much more of an individual and less apparent behaviour in which employees try to establish strict limits to preserve their work-life balance and avoid burnout (Alami et al., 2024; Saraivaa and Nogueirob, 2025). Instead of showing their discontent openly, workers silently stop voluntarily putting any extra effort into their job and limit themselves only to performing formal duties, thereby displaying disengagement without actually leaving their jobs. As stated by Saraivaa and Nogueirob (2025), quiet quitting is a term used to describe the disengagement of employees from their job on an emotional level and their performance of only what is minimally required from them to keep the job, without engaging in activities that go beyond the job description. It is not a quit, but an attempt to minimise their commitment and effort to preserve their work-life balance and avoid burnout, and is exhibited by employees of all generations. Similarly, Alami et al. (2024) stated that quiet quitting refers to a form of silent disengagement where employees reduce their work effort and involvement without formally resigning. It reflects a passive withdrawal characterised by lowered enthusiasm and discretionary effort, often driven by factors like fear of retribution, poor leadership, bureaucratic obstacles, and lack of purpose at work.

Quiet quitting is a concept that is influenced significantly by today's societal and economic forces, including neoliberalism, a philosophy of politics and economy that is characterised by individualism and privatisation as opposed to collectivism (Dillard et al., 2025). The result of this philosophy is increased exploitation of workers, overworking of workers, and mental taxing. Quiet quitting, therefore, is a resistance form that is hyper-individualistic and is meant to counter systemic exploitation, a concept that can be described as a survival tactic. As explained by Dillard et al. (2025), the wider context of quiet quitting is defined by today's societal and economic forces, such as the COVID-19 pandemic, the Great Resignation, economic uncertainty, and inflation, among others. This has led to a change in the way workers look at their work, where the need for autonomy, work-life balance, and fairness increases. This has revealed the disconnects between the employees and the organisational decision-makers, calling for inclusiveness and transparency in workplaces.

**Table 1: Key Differences between Work-to-Rule and Silent Quitting**

| Aspect | Work-to-Rule | Silent Quitting |
|---|---|---|
| Origin | Trade union-led industrial action | Individual coping strategy |
| Objective | Collective protest to improve working conditions | Preserve work–life balance and mental health |
| Visibility | Highly visible and collective | Low-profile and personal |
| Method | Strict rule adherence slows productivity | Withdrawal of discretionary effort |
| Motivation | Bargaining power and labour rights | Burnout prevention, stress management |
| Cultural Context | Unionised, formal labour disputes | Widespread across sectors, including non-union settings |

### 2.2 Contemporary Drivers of Silent Quitting

The modern wave of silent quitting is being shaped by intersecting global, socio-economic, and technological forces. Key global events such as COVID-19, the Great Resignation[1], inflationary pressures[2], and shifting generational values have prompted employees to reassess their relationship with work (Xueyun *et al.*, 2023). Social media platforms, particularly TikTok in 2022, amplified this shift by popularising the idea of doing only what is contractually required to protect mental well-being and avoid burnout (Bhatt *et al.*, 2024; Cashion, 2024; Ellera *et al.*, 2023; Nguyen, 2024). Although the term itself might be contemporary, the behaviours it describes mirror longstanding labour practices and sentiments that workers have exhibited over time. Xueyun et al. (2023) highlighted that the recent surge in discourse on quiet quitting has gained significant global attention through various platforms, demonstrating concerns about employees limiting work efforts to their job requirements to protect their personal lives from workplace stress. They noted that the term, first emerging around 2009, has been popularised recently via social media and sparked social movements and investigations into workplace behaviours and employee well-being. (Bhatt *et al.*, 2024) mentioned that the recent surge in discourse stems from increased awareness of employees deliberately limiting their work engagement to only mandatory duties, as a response to unhealthy workplace environments and unfulfilled expectations. Specifically, in the US, this trend gained prominence through media coverage, including a 2022 Wall Street Journal article[3] highlighting that 50% of employees restrict their job commitment, and social media influencers like Zaid Khan are spreading awareness (Bhatt *et al.*, 2024). The phenomenon reflects growing dissatisfaction among workers who seek better work-life balance without formally resigning, prompting organisations to address underlying causes to retain talent and maintain productivity. Cashion (2024) believed the surge in quiet quitting reflects widespread worker resistance to exploitative labour conditions intensified by contemporary neoliberalism. The study stated that quiet quitting has become relevant as an adaptive approach and a form of hyper-individualised resistance towards neoliberal notions of individualism, which stress personal accountability at the expense of collective effort. The increasing focus on this phenomenon shows the emergence of class awareness and the need for work-life balance in light of declining working conditions and psychological issues. In addition, (Ellera et al., 2023) identified the rise of this phenomenon due to the employee discontent over working in the 21st century, as witnessed by the Great Resignation and the concept of “Quiet Quitters” who confine themselves to doing only what is in their job descriptions.

[1] https://www.bloomberg.com/news/articles/2021-05-10/quit-your-job-how-to-resign-after-covid-pandemic
[2] https://www.investopedia.com/ask/answers/111314/what-causes-inflation-and-does-anyone-gain-it.asp

In the case of African workplaces, silent quitting is further enhanced by various socio-economic factors, including high unemployment rates, economic volatility, and weak labour policies (Utete et al., 2023; Yisa et al., 2024). The workers in these settings experience unstable job security and weak organisational structures that encourage and nurture disengaging behaviours, such as silent quitting, in order to keep their jobs without making any additional effort that will not be appreciated. Yisa et al. (2024) argued that quiet quitting in African workplaces, especially the Nigerian health care sector, is mainly driven by poor remuneration, work overload, poor organisational support, and work-life imbalances. They observed that employees use quiet quitting as one of the strategies to protect themselves from such work stress situations. Utete et al. (2023) argued that there is an urgent need for unionisation of young employees in African workplaces since, presently, union members are mainly between the ages of 45 and 54 years. This means missing out on engaging young employees who have a tendency to quiet quit. They believed this decline in union influence contributes to the rise of quiet quitting, as workers express dissatisfaction through reduced effort and engagement rather than traditional collective action. The findings by Bray (2016) on how happy employees are and what makes some of them think about quitting their jobs, focusing on an audit firm in South Africa, suggested that employees’ uncertainty about pay, promotion, recognition, and working conditions may influence their job satisfaction and intention to leave, which could relate to disengagement behaviours akin to quiet quitting. The study further noted that the cultural and demographic context, such as ethnic group distribution and tenure, also shapes employees' workplace experiences and satisfaction levels. Similarly, Alami et al. (2024) discussed that in African workplaces, employee quiet quitting is influenced by factors such as high bureaucracy, poor working conditions, and leadership challenges. These environments contribute to silent disengagement, where employees quietly reduce effort due to organisational and structural weaknesses, impacting retention and workplace culture. In another similar view, Harris (2024) emphasised employees' generational perspectives and stated that Generation Z[4] employees are particularly prone to quiet quitting due to their unique engagement characteristics and expectations from employers. Additionally, the lack of comprehensive labour laws and enforcement mechanisms exacerbates these challenges, limiting workers' avenues for formal grievances or collective bargaining, thereby pushing many toward subtle forms of withdrawal (Cashion, 2024; Harris, 2024; Utete *et al.*, 2023; Yisa *et al.*, 2024).

[3] https://www.wsj.com/articles/quiet-quitters-make-up-half-the-u-s-workforce-gallup-says-11662517806

[4] https://www.mckinsey.com/featured-insights/mckinsey-explainers/what-is-gen-z

Silent quitting has evolved in a unique manner in terms of the digital work environment, owing to the way technology influences how work is carried out, not only internationally but also in Africa (Adisa et al., 2022; Seeber and Erhardt, 2023). Technology has made it possible for remote work, flexible schedules, and availability at all times. However, it has also made it difficult to distinguish between work and non-work time (Adisa et al., 2022; Seeber and Erhardt, 2023). It has added pressure on employees to be always available, causing them to work even more than stipulated. The resulting increase in the workload and reduction in time available for rest is a primary source of burnout and disengagement among workers, hence quiet quitting. According to Dutta et al. (2024), the extra workload caused by silent quitting is another burden on other employees, as they have to make up for it, thereby increasing their workload. The increased workload, the reduction in downtime, and the lack of rest periods are major contributors to employee burnout and disengagement. In another dimension, Cashion (2024) argued that neoliberal work practices lead to increased workload through unnecessary labour like excessive meetings, causing workers to spend overtime on tasks. This erosion of downtime contributes to burnout, emotional exhaustion, and mental health struggles, as workers feel their self-worth tied to constant productivity but lack control over their time, fostering disengagement and resistance, such as quiet quitting. Similarly, Seeber and Erhardt (2023) highlighted that increased workload and the erosion of downtime, driven by blurred work-home boundaries and frequent digital workplace tool use, contribute significantly to employee burnout and disengagement. These factors strain work-life balance, causing stress and reducing job satisfaction, especially when work permeates personal life without clear boundaries. Additionally, Georgiadou et al. (2025) believed that increased workload combined with the erosion of downtime, especially during phases of company growth or pandemic-induced changes, intensifies job demands while reducing recovery opportunities. This imbalance leads to elevated stress levels, burnout, and emotional disengagement, which in turn foster phenomena like quiet quitting as employees withdraw effort to cope (Bulut *et al.*, 2024; Xueyun *et al.*, 2023).

Early detection and engagement tools, such as AI feedback systems, are essential to prevent silent quitting since they allow organisations to identify any indications of disengagement prior to complete withdrawal (Khan et al., 2025; Madhumita et al., 2024; Sampath et al., 2024). According to Khan et al. (2025), AI-driven feedback systems are able to give real-time insights and monitor the performance of employees to allow proactive approaches to their engagement and assistance. Thus, the creation of a culture of continuous improvement and engagement could help solve issues with quiet quitting and detect disengagement to support and retain employees. It is worth noting that silent quitting usually has its own features that cannot be easily noticed by managers using traditional techniques of observation. AI systems analyse the patterns of communications, performance, and other engagement metrics of employees, detecting the early signs of their disengagement in time. Therefore, such an opportunity to notice problems early and take targeted actions on them, such as providing assistance, redistributing tasks, or training programs for employees, could help resolve the issue of disengagement and reduce the number of silent quits (Madhumita et al., 2024).

Additionally, responsiveness through engagement tools ensures constant open communication between employees and their management, allowing them to share their issues and find ways to solve them. The use of artificial intelligence allows creating mechanisms for the employees to communicate about their experience, requirements, and suggestions in a confidential and easy-to-access way.

According to Ara and Ahmad (2025), transparency through open communication increases employee engagement since the employees feel that they have a say in AI-based HR processes. Such communication ensures that employees feel seen and valued, which is an essential element of their satisfaction and retention with the company. An organisation can decrease the chances of employees choosing to "quietly quit" by providing them with an environment where they are heard and valued. Therefore, by adopting such technology, organisations not only will have opportunities to detect problems in time but also will foster the development of employees' skills and motivation, which is essential for them to remain productive (Madanchian et al., 2023; Weng and Golli, 2024).

Technology has made things even more complicated. Even though the digital age allowed for remote working and hybrid work systems, the lines between work and life have been obliterated (Seeber & Erhardt, 2023; Cashion, 2024). As a result, this culture has overloaded workers and decreased their leisure time, which has contributed to emotional exhaustion and disengagement. The lack of leisure time, in addition to the constant influx of digital communication, makes people feel underappreciated and overburdened. Although a detailed exploration of AI technologies is provided in Section 3, it is worth noting that emerging tools, such as AI-enabled feedback platforms and real-time engagement tracking, are increasingly being positioned as part of the solution. These technologies can help organisations better understand the underlying drivers of disengagement, enabling them to respond proactively with targeted, context-specific interventions.

### 2.3 Psychological and organisational factors contributing to silent quitting

The phenomenon of silent quitting or quiet quitting arises due to a number of psychological and organisational factors leading to employee disengagement and detachment from their work efforts (AL-Jawfi, 2024; Ellera et al., 2023; Gabelaia and Bagociunaite, 2024). From the point of view of psychology, employees suffer from burnout, emotional exhaustion and a sense of losing the meaning of work. Professional burnout acts as an important predictor of the behaviour under consideration, since employees apply it to protect themselves from additional stress and preserve their well-being. The feeling of undervaluation, disrespect, and the failure to satisfy their intrinsic needs causes their dissatisfaction, thus undermining their engagement in work and causing them to focus only on formally required activities.

In terms of organisational issues, ineffective leadership and management are major contributors to quiet quitting. In cases where leaders do not offer appropriate supervisory support, recognition, and participation of employees in decision-making, it is likely to result in disengagement since there is a disconnect between them. For instance, firms which adopt an authoritative style of leadership and communication problems experience frequent cases of quiet quitting (Atalay & Dağıstan, 2024; Esen, 2023). Organisational structures with inadequate reward schemes, no career opportunities, and inconsistency between personal and organisational goals also lead to disengagement. Furthermore, an unhealthy organisational culture characterised by demanding work, low social support, and confusion between work and personal life is a contributing factor to increased stress and dissatisfaction, hence employees choose to quit quietly.

### 2.4 Digital Labour, Workplace Surveillance, and the African Context

To appreciate the concept of silent quitting, it is necessary to place it within the context of the broader process of work transformation amid digital labour regimes. By digital labour, we do not simply mean work that takes place through digital technology; rather, it involves increased mediation of labour processes by data, algorithms, communication technologies and automation of the management process (Alasoini et al., 2023; Wang & Tomassetti, 2024). Particularly, according to Alasoini et al. (2023), digital labour basically means platform work that is managed by smart technology and algorithmic control, and that encompasses "tasks assigned, observed, measured, and assessed through algorithms and ratings." It is seen as a new field of work that differs significantly from the traditional labour markets, and where inequalities are often reproduced and intensified from the offline world. In contemporary workplaces, employees are expected to remain connected via email, messaging platforms, collaboration software, virtual meetings, and digital performance systems that track participation, responsiveness, and output. This shift has altered the conditions under which labour is organised and evaluated, making work more measurable, more visible, and, in many cases, more intrusive.

In such a setting, communication at work is not limited to interpersonal interactions. Instead, it becomes mediated by platforms and data (Treem et al., 2023). Employees' chats, their answers to surveys, task completion rates, meeting participation and log-ins to the system become available to the management and can be used as indicators of productivity, motivation and dedication. The ability of AI-based feedback systems to analyse patterns of communication, sentiment, engagement and behaviour provides an additional layer of managerial information (Kayusi et al., 2025). However, while such systems allow for more prompt reactions and timely interventions, they also challenge the traditional distinction between listening and surveillance at work. Employees see constant monitoring positively when it is associated with being noticed, having the load of their work adjusted and getting assistance. Nevertheless, the same systems can be viewed as coercion if they are used to increase supervision, pressure employees to react and measure their hidden emotional labour (Constantinides and Quercia, 2023; Glavin et al., 2024; Wang et al., 2020).

These tensions are especially relevant when considering the dynamics of digital labour in African workspaces due to their characterisation by uneven infrastructure, sector differences, and more general socio-economic characteristics (Amoako et al., 2023). Access to high-quality internet, electricity, security of devices used, as well as enterprise management systems varies greatly depending on different countries, sectors and organisation types. Employees working for multinationals which are highly digitised may face artificial intelligence dashboards and monitoring of their performance at any moment, whereas workers in governmental bodies, small companies or under-resourced organisations may use fragmented digital platforms or informal channels for communication (Giering and Kirchner, 2025). This uneven development suggests that digital transformation of labour in Africa happens in an inconsistent, uneven and multi-layered way (Chidoori and Van Belle, 2020; Karacuka et al., 2024; Nguimkeu and Okou, 2021). As a result, workers' experiences of monitoring, visibility, autonomy, and feedback vary widely across the continent. Chidoori and Van Belle (2020) pointed out that individuals can be excluded from digital labour markets based on race, gender, or economic status, and that highly-skilled workers may exploit less experienced ones, leading to burdensome, low-paying work and reduced wages for the low-skilled.

Digital labour conditions can also intensify disengagement in less visible ways. The expectation of constant connectivity may expand work beyond formal hours, while digital communication overload can erode downtime and contribute to emotional fatigue (Bondanini, Giovanelli, *et al.*, 2025). Employees may be assessed based on responsiveness measures instead of the quality or difficulty of work they do, and managers may assume that silence, delay in communication, or low engagement online is a poor attitude without understanding the circumstances in which they occur (Bondanini, Sanchez-Gomez, et al., 2025; Kalischko and Riedl, 2021). Where voicing dissent is not encouraged, or there is no strong legislation regarding employee protection, silent quitting may become a way of coping with increased digital pressures without engaging directly in any form of resistance. Therefore, silent quitting in African workplaces should not only be seen as a form of individual disengagement, but as an adaptation to the new structures of digitalised work.

Understanding this context is key to any analysis of AI feedback mechanisms. The AI feedback mechanism is part of wider workplace changes that see data, automation, and digital visibility increasingly becoming part of workplace relations. What makes these AI mechanisms valuable is not only the ability to detect disengagement but also the insights they provide into how organisational structures, communication, and digitalisation affect the experience of workers. But without contextuality and proper governance, the AI feedback tool will misinterpret disengagement, reinforce inequality, and worsen distrust. In the African context, the task is not only to digitalise feedback from workers but to do so, taking into account the local contexts and social environment of digital labour.

## 3. The Role of AI Technologies in Employee Feedback, Engagement, and Labour Management

Artificial Intelligence (AI) technologies have become key instruments in managing employees' experience in modern organisations. AI technologies go beyond HR processes, such as recruitment or automation of administrative functions, and have now become core to everyday labour management using systems that track communication, analyse sentiments, forecast disengagement, and give personalised feedback (Chilunjika et al., 2022; Deranty and Corbin, 2022; Giering and Kirchner, 2025). AI technologies used for giving feedback to employees should be considered not just a mechanism for boosting employee engagement but rather a part of a general trend of digitalised labour management. The development of this trend raises important questions related to how work will be controlled and assessed, and how employees' experience will change in the age of digitally-mediated labour management. This implies an intricate interaction of algorithmic management, individual freedom, and the possibility of reproduction of the same power imbalances between people in organisations (Sorg et al., 2023).

From the point of view of HR, AI provides data-based solutions that can help in making employee feedback processes more objective, fast, and consistent (Madhumita et al., 2024; Sampath et al., 2024). Such tools as natural language processing (NLP), sentiment analysis, predictive analytics, and machine learning models can be used to gather and interpret in real time huge amounts of information related to employees' performance and communication. This technology will allow HR to spot decreasing motivation, recognise the signs of burnout, and take steps before disengagement occurs (Adeusi et al., 2024; Nurjaman, 2025). In this way, the use of AI will be beneficial for organisations that strive to transition from the approach where employee feedback is provided once in a while to more proactive approaches based on continuous engagement. According to Nurjaman (2025), the use of machine learning technologies will be helpful for HR as a means of obtaining insights into workplace culture and employee engagement. Predictive models can identify early warning signs such as decreased engagement or performance drops, allowing HR departments to proactively intervene before employees decide to leave. This enables organisations to address the root causes of turnover and implement targeted interventions to improve retention.

In addition, the increasing use of AI technologies in feedback and performance management is part of a broader trend of workplace automation. AI technologies now play an increasingly mediating role in the dynamic between employees and their managers by interpreting communication patterns, behavioural data, and signs of engagement into insights for the manager. Chatbots, pulse surveys, performance dashboards, and predictive models are not just used to extract data but redefine what participation, responsiveness, and productivity are (Ramu et al., 2026; Wendakoon, 2024). By such technologies, labour increasingly becomes visible in terms of metrics, categorisations, and risk assessments. Even if this can help with making better-informed decisions earlier on, it might also constitute a form of algorithmic surveillance.

For instance, AI-driven sentiment analysis algorithms can analyse employee surveys, internal communications, and feedback portals for signs of frustration, withdrawal, and low engagement (Nurjaman, 2025). Meanwhile, pulse surveys can be used to monitor the changes in morale, while behavioural analysis tools can spot the signs of reduced participation, delays in communications, and decreased collaboration. In this case, these features can allow organisations to deal with silent quitting more efficiently, as managers may often miss the signs of low employee motivation. However, such features can be perceived by employees as intrusive when implemented improperly and poorly governed or when being used as a tool for punishing rather than helping them. In this regard, it should be noted that Adeusi et al. (2024) stressed that these features can be perceived as intrusive by employees unless explained and governed appropriately or unless used for punitive purposes.

Globally, commercial platforms such as Lattice[5], 15Five[6], Workday[7], Degreed[8], Coursera for Business[9], and LinkedIn Learning[10] have demonstrated how AI can support continuous performance review, personalised learning, and engagement analytics (Weng and Golli, 2024). These platforms use machine learning and NLP to automate evaluation tasks, recommend developmental pathways, and generate real-time insights into employee progress. In principle, such tools can improve fairness, reduce managerial blind spots, and help organisations respond more quickly to changing employee needs. In African workplaces, similar technologies may offer significant value where formal feedback systems are underdeveloped or where large, dispersed workforces make continuous engagement difficult (Chilunjika *et al.*, 2022). However, their utility depends heavily on contextual fit, infrastructure, digital literacy, and trust.

In addition, AI could facilitate the expression of voice of employees by developing secure, easy-to-use, and frictionless means of communicating (Gusti et al., 2024). Such applications as anonymous feedback systems, HR chatbots, or even adaptive pulse surveys could encourage employees to raise their voice in cases where employees feel uncomfortable voicing their concerns due to cultural restrictions in extremely hierarchical structures or fear of being penalised for doing so. However, employee trust remains crucial. If workers perceive AI systems as opaque, punitive, or primarily designed to monitor compliance, engagement may decline rather than improve. Thus, the same technologies that promise responsiveness can also reinforce mistrust if not implemented with transparency and care.

[5] https://lattice.com/
[6] https://www.15five.com/
[7] https://www.workday.com/en-za/homepage.html
[8] https://degreed.com/experience/
[9] https://www.coursera.org/business
[10] https://www.linkedin.com/learning/

The use of AI in employee engagement, therefore, presents a dual reality. On one hand, it enables organisations to listen continuously, personalise interventions, and support more adaptive forms of management. On the other hand, it automates aspects of labour governance in ways that can intensify surveillance, narrow managerial interpretation to quantifiable signals, and reduce the space for human judgment. This tension is particularly important in the African context, where organisations operate across widely varying levels of digital maturity and where workers may already face insecurity, limited protections, and unequal access to technological resources. Responsible use of AI must therefore go beyond technical efficiency to include cultural sensitivity, ethical safeguards, and meaningful human oversight.

As a whole, AI tools contribute to a transformation of employee feedback and engagement due to the expanded capabilities of organisations when dealing with gathering, analysing, and using employee data. At the same time, the significance of AI goes far beyond technological innovations since it contributes to changing labour management. Viewing AI as an engagement tool and as a means of exercising control allows for a comprehensive and critical perspective of AI's impact on the modern world of work. The task for African organisations in using AI for feedback is in taking advantage of it while preventing automation from undermining the trust, fairness, and dignity of employees. The need to use AI ethically, which was pointed out by Albu et al. (2025), is essential to keep trust and maintain ethical standards of AI implementation, especially related to fairness and transparency.

### 3.1 Technical Foundations of AI in Employee Engagement

AI-driven employee engagement systems are built on a set of interrelated technologies that enable organisations to collect, interpret, and act on large volumes of employee data in more continuous and sophisticated ways than traditional HR tools. Rather than relying solely on annual appraisals, exit interviews, or occasional satisfaction surveys, these systems draw on computational methods that can identify patterns in communication, behaviour, and performance over time (Kayusi *et al.*, 2025; Nahar *et al.*, 2025). In the context of employee engagement, the technical foundations of AI are especially important because they determine not only what kinds of employee experiences can be measured, but also how disengagement, motivation, and silent quitting are interpreted within organisational systems. The four most significant technical foundations are (1) *Natural Language Processing (NLP)*, (2) *sentiment analysis*, (3) *machine learning*, and (4) *predictive analytics*.

One of the most important enabling technologies used in AI-based engagement systems includes Natural Language Processing (NLP) (Mah et al., 2022; Sharma and Chanana, 2026). The term NLP denotes the capability of machines to process, analyse and interpret human language in such a way that produces insightful information from texts and speeches. At the same time, in the case of workplaces, the use of NLP means the ability of HR systems to analyse texts in which employees provide their feedback in surveys, open questions and chatbots, their internal communications, and descriptions in performance reviews. With NLP, HR managers can categorise topics in employees' feedback and find patterns of employees' experience descriptions without reading every comment by hand (Giering and Kirchner, 2025; Rivera et al., 2021). Thus, the problems related to the work environment, such as workload pressure, absence of recognition, unclear expectations, or distrust in management, can be revealed.

Related closely to NLP is sentiment analysis, which centres on determining the emotional attitude that employees express in their communication. Sentiment analysis is an approach to language classification that can identify whether it is positive, negative, or neutral, while more sophisticated tools for sentiment analysis can detect more subtle emotions, such as frustration, anxiety, dissatisfaction, enthusiasm, or disengagement (Albu et al., 2025; Lee and Song, 2024). When applied to employee engagement software, sentiment analysis becomes especially relevant since silent quitting occurs gradually and starts with changes in tone, language, and communicative energy without making explicit complaints or signalling any intention to leave the job. By analysing these patterns over time, organisations can identify mood trends across teams and intervene before disengagement deepens (Lee and Song, 2024; Pelled and Xin, 1999). However, sentiment analysis must be interpreted carefully, because emotional meaning is highly context-dependent and may vary across cultures, languages, occupations, and communication styles. This is especially relevant in African workplaces, where multilingual environments and culturally specific forms of expression may challenge standard sentiment models trained on non-African datasets.

Machine learning algorithms provide the adaptive capability that allows AI systems to improve their performance over time. Machine learning refers to computational models that learn from historical data in order to identify patterns, classify outcomes, and support decision-making without being explicitly programmed for every scenario (Jordan and Mitchell, 2015). In employee engagement contexts, machine learning systems may be trained on past data relating to absenteeism, survey responses, communication activity, performance ratings, training participation, turnover history, or feedback frequency. By learning from these variables, the system can identify combinations of indicators associated with disengagement, burnout, declining productivity, or eventual attrition. For instance, a model might detect that reduced participation in team communication, declining survey positivity, and repeated overtime patterns often precede lower engagement or resignation risk. These insights can help HR teams and managers identify employees or teams that require attention, support, or workload adjustments. The value of machine learning lies in its ability to reveal patterns that may not be easily visible through conventional observation alone (Adeusi *et al.*, 2024; Alami *et al.*, 2024; Nurjaman, 2025).

Predictive analytics leverages both statistical modelling and machine learning to forecast future workforce outcomes from current and historical data (Kakulapati *et al.*, 2020; Ramu *et al.*, 2026). In employee engagement systems, predictive analytics is used to estimate risks such as declining job satisfaction, absenteeism, low morale, reduced discretionary effort, or employee turnover (Nahar *et al.*, 2025). These models allow organisations to shift from reactive HR management to more proactive intervention by identifying likely future problems before they become organisational crises. For example, predictive models may estimate which employee groups are most vulnerable to burnout, which units are showing signs of disengagement, or where retention challenges are likely to emerge. In relation to silent quitting, predictive analytics is particularly useful because it captures the gradual and often hidden progression from dissatisfaction to emotional withdrawal. Rather than waiting until productivity falls sharply or employees resign, organisations can use predictive indicators to initiate mentoring, support discussions, redesign roles, or implement well-being interventions at an earlier stage. This makes predictive analytics a strategic tool for workforce stability and employee support.

Overall, these technical foundations enable AI-driven engagement systems to function as continuous listening mechanisms within the organisation. NLP makes employee language visible as data, sentiment analysis interprets emotional tone, machine learning identifies hidden patterns, and predictive analytics forecasts likely outcomes. Their combined use allows organisations to transform fragmented feedback into actionable insight. However, the power of these technologies also raises important concerns. The quality of outputs depends heavily on the quality of the data, the assumptions embedded in the model, and the context in which the technology is deployed (Adeusi *et al.*, 2024; Nurjaman, 2025). Poorly designed systems may misclassify emotions, exaggerate risk, or lead to misleading conclusions about employee motivation. Similarly, overreliance on automated interpretation may reduce opportunities for contextual explanation and human judgment.

For African workplaces, the technical foundations of AI in employee engagement must be understood within conditions of uneven digital maturity, linguistic diversity, and infrastructural variability. Models developed in highly digitised Western contexts may not perform effectively in workplaces characterised by multilingual communication, informal management structures, variable internet access, and different cultural norms around expression and authority (Chidoori and Van Belle, 2020; Chukwuka and Dibie, 2024; Kayusi *et al.*, 2025). As a result, the technical robustness of AI systems in African settings depends not only on computational accuracy, but also on contextual adaptation. Organisations must therefore ensure that AI models are trained, tested, and governed in ways that reflect local realities rather than assuming universal applicability.

Overall, the technical foundations of AI in employee engagement provide the analytical engine that enables organisations to monitor employee experience, identify disengagement, and support more timely interventions. Yet these technologies should not be seen as neutral or self-sufficient. Their effectiveness depends on ethical design, cultural sensitivity, and human oversight. When deployed responsibly, they can help organisations better understand silent quitting and improve engagement. When deployed uncritically, they risk reducing complex human experiences to simplistic signals and reinforcing the very disconnection they are meant to address.

### 3.2 AI-Powered Feedback Tools and Mechanisms

The practical value of AI in employee engagement becomes most visible through the tools and mechanisms organisations use to capture, interpret, and respond to employee experience in real time. While the technical foundations discussed in the previous subsection explain how AI systems process data, this subsection focuses on the concrete applications through which those capabilities are operationalised in workplaces. These tools are designed to strengthen communication, improve responsiveness, and reduce the time lag between employee dissatisfaction and managerial intervention. In the context of silent quitting, their importance lies in helping organisations detect subtle changes in employee behaviour before disengagement becomes deeply embedded. Among the most widely used mechanisms are *AI-powered pulse surveys (sample in Figure 1)*, *HR support chatbots*, *engagement dashboards*, and *behavioural analytics tools*.

AI-powered pulse surveys are one of the most accessible and widely adopted feedback mechanisms in contemporary HR practice. Unlike traditional annual engagement surveys, which often yield delayed, overly general results, pulse surveys are short, frequent, and designed to capture employee sentiment on an ongoing basis (Brown, 2022; Klinger *et al.*, 2026). Their AI-enhanced value lies in their adaptability and analytical depth. These systems can adjust questions based on previous responses, automatically identify emerging themes, and detect patterns across teams, roles, or time periods. For example, if repeated responses indicate concerns about workload, leadership communication, or work-life balance, the system can generate follow-up prompts that explore these issues more deeply (Malik *et al.*, 2023). This dynamic design allows organisations to move beyond static measurement toward a more continuous listening approach. In relation to silent quitting, pulse surveys are particularly useful because they can reveal declining morale, reduced trust, or increasing emotional detachment long before these issues become obvious through formal performance indicators.

How was your last month at work?

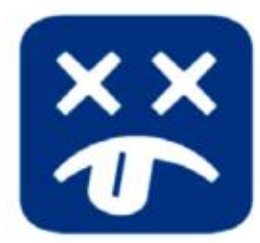
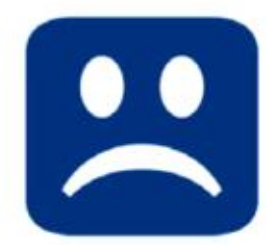
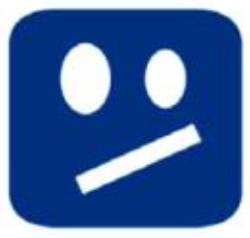
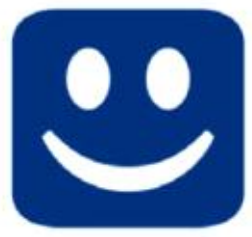
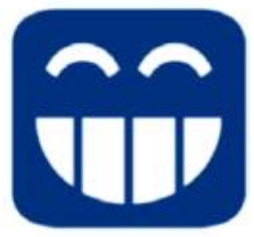

**Figure 1. A Typical example of pulse surveys**

**Source: (Brown, 2022)**

Another important tool is AI-driven chatbots for HR support. These conversational tools provide employees with an immediate and interactive channel for asking questions, reporting concerns, and accessing workplace information (Malik *et al.*, 2022). HR chatbots can assist with routine matters such as leave policies, benefits, onboarding, performance procedures, and training opportunities, thereby reducing administrative delays and improving access to information. More importantly, they can also function as informal feedback channels by collecting employee concerns in real time and identifying recurring patterns in the topics employees raise. In hierarchical or highly formal workplaces, chatbots may offer employees a less intimidating avenue for expression than direct interaction with managers or HR officers. This may be especially relevant in African organisational contexts where cultural norms, power distance, or fear of reprisal discourage open communication. However, the effectiveness of chatbots depends on trust and usability (Malik *et al.*, 2022; Pillai *et al.*, 2024). Employees are more likely to engage with these tools when they believe their concerns will be treated confidentially, interpreted fairly, and followed by meaningful action.

Engagement dashboards represent another major application of AI in employee feedback systems. These dashboards are visual interfaces that aggregate and present employee engagement data in ways that support managerial awareness and decision-making (Ramu *et al.*, 2026; Wendakoon, 2024). They may display trends such as survey participation rates, sentiment shifts, communication levels, absenteeism patterns, recognition frequency, or turnover risk indicators across different teams or departments. By converting complex datasets into accessible visual summaries, dashboards enable managers and HR practitioners to monitor organisational climate more continuously and identify areas requiring attention (Richardson *et al.*, 2025). In the context of silent quitting, dashboards can be particularly valuable because they make disengagement patterns visible at both individual and collective levels. A team showing consistently lower sentiment, declining participation, or reduced feedback engagement may require intervention even if formal performance outcomes have not yet deteriorated. However, dashboards also shape what managers notice and prioritise. If poorly designed, they may encourage an overly metric-driven understanding of engagement, reducing rich human experiences to simplified indicators (Wendakoon, 2024). Their usefulness, therefore, depends on thoughtful interpretation and complementary human judgment.

Behavioural analytics tools extend the reach of AI-driven engagement systems by analysing patterns of workplace activity and interaction (Kakulapati *et al.*, 2020; Sharma and Chanana, 2026). These tools monitor indicators such as communication frequency, meeting participation, task completion timing, collaboration patterns, digital responsiveness, and workflow engagement. Their purpose is to identify changes in behaviour that may signal stress, withdrawal, overload, or declining commitment (Nahar *et al.*, 2025). For example, a sharp drop in participation in team platforms, reduced responsiveness to collaborative tasks, or significant shifts in work rhythm may suggest emerging disengagement. In this sense, behavioural analytics tools are often used as early warning systems, particularly in remote or hybrid workplaces where managers have fewer opportunities for direct observation (Boccoli *et al.*, 2023). In African workplaces, where remote work adoption is uneven but growing, such tools may help organisations understand how digital work conditions are affecting employee connection and contribution. At the same time, they raise significant concerns about surveillance, privacy, and the potential misinterpretation of context. A reduction in online activity, for instance, may reflect infrastructure problems, caregiving burdens, or digital fatigue rather than intentional disengagement. This means behavioural data must be interpreted cautiously and ethically.

Taken together, these AI-powered tools and mechanisms enable organisations to shift from episodic engagement management to more continuous, data-informed approaches. Pulse surveys capture evolving employee attitudes, chatbots facilitate immediate communication, dashboards support strategic visibility, and behavioural analytics tools detect patterns that might otherwise go unnoticed. Their combined use can significantly improve organisational responsiveness, particularly where silent quitting emerges gradually and without open confrontation. These mechanisms also have the potential to strengthen employee voice by creating multiple channels through which concerns, frustrations, and support needs can be recognised.

### 3.4 Global Case Examples

The examples used here demonstrate how AI-driven feedback systems are being applied across large, complex organisations to improve employee listening, internal support, and managerial responsiveness. These examples are useful because they show that AI in employee engagement is not limited to a single tool or organisational model. Instead, organisations are combining employee listening platforms, AI-supported service tools, and generative AI summarisation to strengthen feedback loops and improve how leaders respond to workforce needs. At the same time, these examples show that the value of such systems depends on how well they are integrated into broader organisational processes rather than treated as standalone technologies.

A useful example is *Delivery Hero*[11], a global delivery platform, which *Workday* describes as using *Workday Peakon Employee Voice*[12] to understand the daily experience and engagement levels of more than 17,000 employees worldwide (Workday, 2026a). According to Workday's customer case, the company needed a system that could collect regular employee feedback in multiple languages and provide insight across the business. This example is particularly relevant because it illustrates the role of AI-enabled employee listening in a geographically dispersed workforce, where managers may otherwise struggle to identify changes in morale, participation, or engagement across teams operating in different countries and cultural settings (Giering and Kirchner, 2025). In the context of this paper, *Delivery Hero* represents a global case of AI-supported engagement monitoring designed to make employee experience more visible and actionable at scale.

[11] https://www.deliveryhero.com/

[12] https://www.workday.com/en-us/products/employee-voice/overview.html

A second example is *Microsoft*, which has publicly documented its use of AI-powered internal HR support tools (Microsoft, 2026a). Microsoft reports that its HR organisation adopted Dynamics 365 Customer Service with Copilot[13] to modernise employee support, introducing AI-powered case summaries, automated email drafting, and knowledge assistance for HR advisors. Microsoft says this helped its HR function deliver faster and more responsive support for employees while reducing friction associated with legacy systems and manual case handling (Microsoft, 2026b). This example is important because it shows that AI in employee engagement is not only about sentiment analysis or surveys; it also includes internal service automation that affects how employees experience organisational support. In practical terms, when routine HR interactions become faster, clearer, and easier to access, organisations may improve employee experience and reduce some of the frustrations that contribute to disengagement.

A third example is *Johnson & Johnson*, which has been publicly associated with the use of employee listening and generative AI to improve employee experience, well-being, and development (Zielinski, 2024). Qualtrics highlighted transforming employee listening through experience management and generative AI, and external reporting (Herbert, 2024). *Johnson & Johnson* used generative AI to summarise large volumes of employee survey comments and provide more personalised feedback from engagement data (Qualtrics, 2026). This example is especially relevant because it illustrates how AI can be used not only to collect employee voice data but also to interpret large-scale qualitative feedback in ways that are more manageable for leaders and more meaningful for employees. In the context of silent quitting, this kind of application is important because disengagement often emerges first in comments, tone, and narrative feedback rather than in formal performance indicators alone (Madhumita *et al.*, 2024).

Taken together, these global examples reveal three important patterns. First, AI-driven engagement systems are being used in different ways: to capture multilingual employee feedback across borders, to improve internal HR support, and to summarise complex employee sentiment for managerial action. Second, these systems are most valuable when they help organisations move from periodic, reactive feedback processes to more continuous and responsive models of engagement. Third, the examples also suggest that AI is increasingly becoming part of the wider infrastructure of labour management, shaping how employee voice is collected, interpreted, and acted upon in digitally mediated workplaces. These global cases show that AI-driven feedback systems can improve organisational listening and support earlier intervention when employee disengagement begins to develop. However, they also reinforce a key point developed throughout this paper: the effectiveness of such systems depends on human oversight, contextual interpretation, and organisational trust. For this reason, while global cases provide useful models of technological possibility, their relevance to African workplaces lies not in direct imitation but in adapting their lessons to local labour conditions, cultural contexts, and institutional realities.

[13] https://www.microsoft.com/en-us/microsoft-copilot/copilot-101/ai-for-hr

## 4. Case Studies from African Organisations

This section uses publicly documented examples from African organisations to illustrate how AI-enabled systems, digital HR platforms, and internal automation tools are reshaping workplace management across the continent. These examples do not all represent identical kinds of AI deployment, nor do they all focus narrowly on employee engagement. Rather, they show different but related dimensions of workplace transformation, including integrated human capital management, AI-assisted productivity, internal support automation, and data-driven HR administration. They demonstrate that the transformation of African workplaces is already underway, although its pace, depth, and organisational logic vary significantly across sectors.

### 4.1 Absa Group: Integrated HR Systems and Faster Talent Review

*Absa Group*[14] provides one of the clearest publicly documented examples of large-scale HR digital transformation in Africa (Workday, 2026b). According to *Workday's* customer case material, Absa adopted Workday's end-to-end HR solution to unify people processes across headquarters and operations in nine African countries. The same source states that, before this transition, the bank was operating 25 separate systems and that fewer than 10% of management roles were covered by succession planning. Workday further reports that Absa reduced talent review times from up to 10 days to four hours, enabling much broader review coverage across the organisation.

From this paper's perspective, Absa is relevant because it demonstrates how digital HR infrastructure can reshape labour management at scale. A system of this kind does not merely digitise administration; it changes how performance, succession, workforce visibility, and managerial decision-making are organised. It also illustrates how digital platforms can increase organisational capacity to monitor talent flows and standardise people processes across multiple national contexts. While this is not the same as a narrowly defined employee feedback system, it clearly demonstrates the transition toward more data-driven, automated workplace governance in an African organisation.

Absa also appears in Microsoft customer materials on Copilot, where the company is described as using AI-driven tools to enhance efficiency across business areas and reduce employees' administrative burden (Microsoft, 2024a). This is useful for this paper because it shows a second layer of workplace transformation: beyond HR systems, AI is also entering day-to-day work practices and employee productivity routines.

### 4.2 Standard Bank: Enterprise Automation, HR Applications, and Generative AI Readiness

*Standard Bank*[15] offers another important African example of workplace transformation through digital systems. Microsoft reports that the bank has developed more than 1,500 Power Platform solutions across its global network, including applications for finance, operations, and HR (Microsoft,

[14] https://www.absa.africa/about-us/
[15] https://www.standardbank.co.za/southafrica/personal

2024b). The same source notes that Standard Bank supports more than 1,500 citizen developers and is exploring generative AI capabilities through Microsoft Copilot Studio and related tools.

This case is significant because it illustrates that digital labour transformation in African workplaces is not limited to a single HR platform or AI feature. Instead, it can emerge through a broader organisational ecosystem in which employees, managers, and technical teams build and use internal digital tools that shape workflows, approvals, reporting structures, and people-management processes. In this sense, Standard Bank demonstrates how enterprise automation can gradually reshape labour processes, including HR functions, without necessarily presenting itself as a standalone employee-engagement initiative.

For this paper, Standard Bank is especially relevant as an example of how large African firms are building internal digital capacity that may later support more advanced forms of AI-enabled engagement monitoring, internal communication analysis, and managerial decision support. Its example, therefore, fits the broader theme of digital labour and automation by showing how organisational infrastructures for AI adoption are being assembled in African workplaces.

### 4.3 Safaricom: Internal Support Automation and AI-Enabled Employee Assistance

*Safaricom*[16] provides a useful Kenyan example of internal AI-enabled workplace support. Safaricom publicly operates *Abby*[17], which it describes as an internal support assistant for employees and vendors, available as a virtual IT assistant. In addition, Safaricom has publicly stated that it has been using AI and machine learning for more than 5 years across various parts of its business, and its newsroom has reported that all its employees have become certified in AI and other emerging technologies (Safaricom, 2023).

The value of Safaricom as a case study lies in its integration of AI not only into customer-facing innovation but also into internal workplace support and workforce capability-building. An internal assistant such as Abby may not be a full employee-sentiment system, but it does represent an important form of AI-enabled labour support: employees can access assistance, information, and routine support through a digital interface rather than depending solely on manual internal service channels. This matters for workplace transformation because it changes how workers interact with organisational systems and how support functions are delivered at scale.

Safaricom is also useful because it shows that workforce transformation involves both skills and systems. Public reporting on company-wide AI certification suggests an effort to prepare employees for a workplace in which AI tools are increasingly embedded in daily work, decision-making, and service delivery (Safaricom, 2025). In the context of this paper, Safaricom therefore represents a case where internal support automation and employee upskilling intersect in ways that are relevant to the future of work in Africa.

[16] https://www.safaricom.co.ke/
[17] https://abbybot.safaricom.co.ke/

### 4.4 Adcorp: Consolidating Fragmented Human Capital Systems

*Adcorp*[18] offers a South African example of HR system consolidation as a foundation for more data-driven labour management. Workday's case material states that Adcorp selected Workday to integrate previously fragmented HR, finance, and operational systems, reducing its HCM systems from four to one (Workday, 2026c). The same material reports that standardising processes, including leave rules, generated major operational efficiencies and improved data intelligence across the business.

Although Adcorp's published case is framed primarily in terms of integration and efficiency rather than AI-powered engagement, it remains relevant to this paper because integrated HCM architecture is often a prerequisite for more advanced analytics, automation, and employee-experience management. In practical terms, organisations cannot easily implement intelligent feedback systems, predictive retention tools, or workforce dashboards when their people data is fragmented across disconnected systems. Adcorp therefore illustrates a foundational stage in workplace transformation: building the digital infrastructure through which more advanced labour-management technologies can later operate (Malik *et al.*, 2023).

### 4.5 Emerging Contexts: Egypt, Nigeria, Morocco, and Other African Markets

Beyond South Africa and Kenya, publicly documented African examples of AI-enabled workplace transformation are emerging, though the evidence is uneven in depth and often stronger on digital transformation and productivity than on employee feedback systems specifically. Nigeria currently provides some of the clearest additional organisational examples, while Egypt and Morocco are better understood as developing contexts in which AI adoption, digital capability-building, and workplace modernisation are becoming more visible (Microsoft, 2024c; SAP News, 2025). A broader pan-African layer is also taking shape through infrastructure and ecosystem investments that may enable more advanced workplace AI deployment across multiple countries in the near future.

In Nigeria, one notable example is *Food Concepts Plc*[19], which SAP reported in 2025, had chosen SAP *SuccessFactors* to transform its human capital management capabilities across its West African operations (SAP News, 2025). SAP's account presents the initiative as an effort to strengthen talent attraction, retention, and employee experience, making it one of the more directly relevant public examples for this paper regarding labour management and workplace transformation. Although this case is framed by the vendor and therefore should be interpreted cautiously, it is still useful because it shows a named Nigerian company publicly linking digital HR transformation to workforce management and employee experience.

A second Nigerian example is *Access Holdings Plc*[20], which Microsoft describes as using Copilot for Microsoft 365 to improve work around data management, meeting productivity, and software development (Microsoft, 2024c). Microsoft's customer story says employees became more engaged in meetings and were able to analyse data faster using the tool. This example is not a dedicated HR feedback case, but it remains highly relevant to this paper because it illustrates a different dimension of digital labour transformation: AI entering everyday knowledge work and reshaping how employees collaborate, prepare, code, analyse information, and participate in meetings. In that sense, Access Holdings reflects the spread of AI into the day-to-day labour process rather than into HR systems alone.

[18] https://www.adcorpgroup.com/
[19] https://foodconceptsplc.com/company/
[20] https://theaccesscorporation.com/about/

In Egypt, public evidence is more visible in broader organisational AI adoption than in employee engagement systems specifically. One commonly cited case is *Commercial International Bank (CIB Egypt)*[21] and its Arabic-speaking AI chatbot *Zaki*, which was described as a significant banking AI deployment in Egypt (CIB News, 2024). Publicly available material frames *Zaki* primarily as a customer-facing service innovation rather than an internal employee feedback or HR tool. Even so, the case is still useful in this paper as an indicator of how major Egyptian organisations are participating in AI-led digital transformation and building familiarity with AI-enabled service systems. It therefore represents an emerging context rather than a direct employee-engagement case.

For Morocco, the publicly documented concept is stronger in terms of ecosystem development, AI skills development, and infrastructure growth than in named firm-specific employee feedback systems. Oracle announced expanded research and development investment in Morocco, including work using cloud, AI, and machine learning technologies from its Morocco Development Centre in Casablanca (Oracle, 2025). In addition, public reporting on the *Holmarcom–Microsoft* AI Institute points to growing efforts to build local AI capability through structured training and hybrid learning models (7newMorocco, 2024). These developments do not constitute direct workplace-engagement case studies, but they do indicate that Moroccan organisations and institutions are preparing the skills and technical environment required for broader AI-enabled workplace transformation.

A broader pan-African infrastructure example is provided by *Cassava Technologies*[22], which announced plans to deploy NVIDIA accelerated computing and AI software in South Africa, with expansion planned at its data centre facilities in Egypt, Kenya, Morocco, and Nigeria (CassavaTechnologies, 2026). Cassava describes this as building an African AI Factory that will provide AI-as-a-Service through its continental network. While this is not an HR case, it is highly relevant to the future of digital labour in Africa because workplace AI adoption depends not only on organisational interest but also on infrastructure, compute access, and regional digital ecosystems (Chilunjika *et al.*, 2022; Sharma and Chanana, 2026). The significance of this example is that it suggests the next wave of AI-enabled workplace transformation in Africa may be shaped not only by firms adopting tools, but also by the emergence of continental AI infrastructure that lowers the barriers to adoption across multiple countries.

These emerging contexts show that AI-enabled workplace transformation across Africa is spreading through different pathways. In some cases, such as *Food Concepts* in Nigeria, the entry point is explicitly through human capital management and employee experience. In others, such as *Access Holdings* or *CIB Egypt*, the entry point is broader organisational productivity or AI-enabled service systems. In Morocco and in cross-border infrastructure initiatives such as *Cassava's*, the emphasis is more on readiness, capability-building, and digital ecosystem development. These examples suggest that African workplace transformation is broadening geographically, but the public evidence base remains uneven. For that reason, it is important to distinguish between directly documented employee-engagement systems and broader organisational AI adoption when analysing the relationship between digital labour, automation, and silent quitting across the continent.

[21] https://www.cibeg.com/en/about-us

[22] https://www.cassavatechnologies.com/about-us/

## 5. Challenges: Ethical, Cultural, Technical, and Socioeconomic Constraints

While AI-driven feedback systems offer important opportunities to strengthen employee engagement and identify early signs of silent quitting, their adoption also introduces complex challenges. These challenges are not only technical, but also ethical, cultural, and socioeconomic. In African workplaces, where digital infrastructure, labour protections, organisational cultures, and access to technological resources vary widely, these constraints are especially significant. If not addressed carefully, AI systems intended to improve employee well-being may instead deepen mistrust, reinforce inequality, and create new tensions between workers and employers (Ara and Ahmad, 2025). A responsible approach, therefore, requires organisations to understand the wider implications of AI adoption rather than viewing these systems as neutral tools of efficiency.

### 5.1 Ethical Challenges

One of the most pressing concerns in the use of AI-driven feedback systems is the ethical tension between supporting employees and monitoring them. Organisations have a legitimate interest in identifying disengagement early, improving communication, and reducing the risks associated with silent quitting. However, this objective must be balanced against employees' rights to autonomy, dignity, and privacy (Nahar *et al.*, 2025). When AI systems are used to analyse employee communication, track behavioural patterns, or infer emotional states, workers may feel that the boundaries between legitimate support and intrusive monitoring have become blurred. This is particularly problematic where data is collected continuously or without sufficient explanation, as employees may come to perceive engagement systems as forms of hidden surveillance rather than mechanisms for care and support (Alasoini *et al.*, 2023; Ara and Ahmad, 2025).

Fairness is another major ethical issue. AI-driven HR systems often appear objective because they rely on data, algorithms, and automated analysis (Weng and Golli, 2024). Yet these systems are shaped by the assumptions built into their design, the quality of the data on which they are trained, and the organisational purposes for which they are used. If historical workplace data reflects unequal treatment, managerial bias, or structural exclusion, AI systems may reproduce or even amplify those inequities (Nurjaman, 2025). For example, employees who communicate differently because of cultural background, language style, disability, or role-specific work patterns may be misclassified as disengaged or underperforming. In such cases, the system's appearance of neutrality may conceal unfair outcomes, making ethical oversight essential.

Privacy is equally central to the ethical debate (Sharma and Chanana, 2026). AI-driven feedback systems often rely on analysing large volumes of employee data, including survey responses, communication content, behavioural indicators, and, sometimes, digital interaction patterns. Even when data is collected for legitimate organisational purposes, employees may reasonably question how much information should be gathered, who has access to it, how long it is retained, and whether it may later be used for disciplinary decisions (Chukwuka and Dibie, 2024). These concerns are intensified in contexts where data protection policies are weak, organisational communication is limited, or employees have little confidence in existing grievance mechanisms (Adeusi *et al.*, 2024; Nurjaman, 2025). Ethical AI implementation, therefore, requires clear boundaries around data collection, strong consent procedures, and robust governance on how employee data is interpreted and applied.

Transparency also remains a critical ethical requirement. Employees are more likely to accept AI systems when they understand what the system does, what kinds of data it uses, and how its outputs will influence managerial decisions (Böhmer and Schinnenburg, 2023; Fenwick *et al.*, 2024). Opaque systems create suspicion, especially when workers are unaware that their communication patterns, survey comments, or participation rates are feeding into engagement analytics. Ethical practice, therefore, demands that organisations communicate clearly about the purpose, scope, and limitations of AI-driven feedback systems. Such transparency is not simply a technical matter; it is part of building legitimacy and trust.

Ultimately, the ethical challenge lies in ensuring that AI systems remain aligned with human-centred values. If AI is used primarily to discipline, rank, or pressure workers into visible compliance, it may intensify disengagement rather than reduce it. By contrast, when it is governed transparently and used to support employee well-being, fair treatment, and meaningful responsiveness, it can contribute positively to workplace transformation. Ethical governance is therefore not an optional addition to AI implementation; it is a foundational condition for its legitimacy (Ara and Ahmad, 2025; Madanchian *et al.*, 2023; Weng and Golli, 2024).

**5.2 Cultural Challenges**

AI-driven feedback systems also face significant cultural challenges, especially when deployed across diverse organisational and national settings (Malik *et al.*, 2022). Silent quitting is not experienced, expressed, or interpreted in the same way everywhere. Employee disengagement is shaped by local workplace norms, power relations, generational expectations, communication styles, and broader social values (Ara and Ahmad, 2025; Georgiadou *et al.*, 2025). In some contexts, reduced discretionary effort may be interpreted as a sign of burnout and boundary-setting, while in others it may be judged more harshly as disloyalty or poor work ethic. These differences mean that organisations cannot assume that a single model of engagement, motivation, or feedback will apply uniformly across all employees or workplaces.

In African contexts, cultural diversity adds an additional layer of complexity. Many organisations operate in multilingual environments and bring together workers from different ethnic, national, and social backgrounds (Chidoori and Van Belle, 2020). Forms of communication that appear restrained, indirect, or deferential may be culturally normal rather than signs of withdrawal. Likewise, employees may avoid open criticism of supervisors not because they are satisfied, but because respect for authority, fear of reprisal, or organisational culture discourages direct expression. AI systems trained in different cultural settings may misinterpret these patterns if they are not carefully adapted. This creates a risk that disengagement models built elsewhere may fail to capture the real meanings of silence, politeness, indirectness, or reduced visibility in African workplaces.

Generational diversity also matters. Younger employees may be more comfortable with digital communication, flexible work arrangements, and frequent feedback systems, while older workers may have different expectations regarding authority, communication channels, and performance evaluation (Xueyun *et al.*, 2023). Perceptions of AI may therefore vary significantly within the same organisation. Some employees may see AI-powered feedback tools as modern, accessible, and empowering, while others may experience them as impersonal, disruptive, or threatening. Organisations must take these differences seriously if they want digital engagement systems to be widely accepted and meaningfully used.

Another cultural challenge involves the transfer of workplace technologies across contexts without sufficient localization (Chidoori and Van Belle, 2020; Chilunjika *et al.*, 2022; Malik *et al.*, 2022). Many AI tools used in HR have been designed in highly digitised settings and reflect assumptions about language, organisational structure, employee voice, and individual behaviour that may not hold in all African workplaces. Where imported systems are used without contextual adaptation, they may fail to identify genuine concerns or produce misleading conclusions about employee sentiment and engagement. Tailored implementation is therefore essential. This includes adapting language models, reviewing cultural assumptions in survey instruments, and ensuring that intervention strategies reflect local workplace realities rather than generic global templates.

In this sense, the cultural challenge is not only about diversity; it is about relevance. AI systems can only contribute meaningfully to engagement and disengagement management when they align with how workers actually communicate, interpret authority, and experience work in their own contexts. Cultural sensitivity is therefore central to both technical performance and organisational legitimacy.

### 5.3 Technical Challenges

The technical challenges associated with AI-driven feedback systems are substantial and can significantly affect their usefulness and credibility (Chilunjika *et al.*, 2022; Madhumita *et al.*, 2024). Although these technologies promise greater efficiency, accuracy, and responsiveness, their performance depends on the quality of the underlying data, the appropriateness of the model design, and the organisational capacity to interpret outputs responsibly. Poorly implemented AI systems may misread employee behaviour, exaggerate disengagement risks, or generate misleading insights that create more confusion than clarity (Malik *et al.*, 2022).

One important challenge is the difficulty of detecting complex human experiences through data-driven systems. Silent quitting is not a single, easily measurable event; it is a gradual and context-dependent process shaped by emotion, motivation, workload, leadership, and social relations. AI systems may identify proxies for disengagement, such as reduced communication, lower survey positivity, or changes in collaboration patterns, but these proxies are not always reliable on their own. A drop in responsiveness, for example, may reflect burnout, poor connectivity, family responsibilities, or infrastructural barriers rather than intentional withdrawal. This means that the technical capacity to detect patterns must always be accompanied by careful contextual interpretation (Malik *et al.*, 2023).

Algorithmic bias is another major technical concern. AI systems are only as fair and accurate as the data and assumptions on which they are built (Madhumita *et al.*, 2024). If datasets are incomplete, unrepresentative, or historically biased, model outputs may distort rather than clarify employee experience (Malik *et al.*, 2022). This is particularly important in African workplaces, where digital records may be inconsistent, multilingual data may be difficult to process accurately, and informal work practices may not fit neatly into standardised data structures. Systems designed for other labour markets may perform poorly if applied without retraining, local testing, or linguistic adaptation. Technical robustness, therefore, requires more than adopting an existing platform; it requires validating whether the system actually works under local organisational conditions.

A further challenge involves transparency and explainability. Many AI tools generate scores, classifications, or predictions that may influence managerial decisions, yet the logic behind those outputs is not always visible to users (Kayusi *et al.*, 2025; Weng and Golli, 2024). If managers cannot explain why a system has flagged a team as disengaged or labelled an employee as high risk, they may either overtrust the output or ignore it altogether. Employees, in turn, may lose trust in systems whose judgments appear arbitrary or opaque. Explainability is, therefore, not only an ethical issue but also a technical one, because systems that cannot be meaningfully interpreted are less likely to be used responsibly (Madhumita *et al.*, 2024).

The reduction of human oversight is also a technical risk. AI systems can support faster decision-making, but when organisations become overly reliant on automated outputs, the space for managerial judgment, employee explanation, and contextual correction may shrink (Weng and Golli, 2024). This can be especially harmful in high-impact HR situations, where individual circumstances matter greatly. Maintaining human involvement in reviewing AI-generated insights is therefore essential to ensuring that automated systems remain tools of support rather than substitutes for thoughtful management (Kayusi *et al.*, 2025).

Finally, technical integration remains a practical challenge. Many organisations, especially in resource-constrained settings, operate with fragmented HR systems, incomplete employee records, limited analytics capacity, and weak interoperability between platforms. Without reliable infrastructure and strong internal capability, even sophisticated AI tools may fail to deliver meaningful benefits (Malik *et al.*, 2023). Technical readiness is, therefore, a major determinant of whether AI-driven feedback systems can function effectively in practice.

### 5.4 Socioeconomic Impacts of AI-Driven Labour Management in African Workplaces

Beyond ethical, cultural, and technical concerns, AI-driven feedback systems also have broader socioeconomic implications for African workplaces. As organisations automate aspects of labour management through sentiment tracking, performance analytics, behavioural monitoring, and predictive feedback systems, the structure and quality of work may change in ways that extend far beyond HR efficiency (Kayusi *et al.*, 2025). These technologies influence how labour is valued, how productivity is measured, and how workers experience autonomy, pressure, and organisational belonging. In this sense, AI-driven feedback systems are not merely administrative innovations; they are part of a wider transformation in the political economy of work across digitally evolving African organisations.

One important implication is the intensification of performance visibility. AI systems enable managers to monitor participation, responsiveness, communication patterns, and output with greater speed and granularity than traditional supervisory methods (Ara and Ahmad, 2025; Madanchian *et al.*, 2023). While this may improve organisational responsiveness, it can also deepen pressure on employees to remain constantly available, digitally visible, and behaviourally compliant. In workplaces already marked by high unemployment and precarious job security, workers may feel compelled to perform engagement rather than genuinely experience it. This can create a paradox in which digital feedback tools are introduced to improve employee well-being, yet simultaneously increase anxiety, self-monitoring, and emotional strain (Bondanini, Giovanelli, *et al.*, 2025).

AI-driven labour management may also widen inequalities between workers and organisations, as well as among workers themselves. Employees in highly digitised environments may have greater access to learning platforms, personalised development pathways, and structured feedback, while those in under-resourced workplaces may encounter fragmented systems, inadequate support, or exclusion from digital opportunities altogether (Chidoori and Van Belle, 2020; Madhumita *et al.*, 2024). Differences in digital literacy, device access, connectivity, and infrastructure can shape who benefits from AI-enhanced management and who becomes more vulnerable under it. These patterns are especially significant in African contexts, where technological access and organisational capacity vary sharply across sectors, regions, and institutional types. Without deliberate inclusion strategies, AI adoption may reinforce existing inequalities rather than reduce them.

Another socioeconomic concern is the changing nature of managerial power. AI systems often shift decision-making toward data-driven classifications, engagement scores, and predictive indicators, which may appear objective but remain shaped by the assumptions embedded in the model design (Weng and Golli, 2024). When organisations rely too heavily on automated interpretation, workers may lose opportunities to explain context, challenge evaluation, or negotiate expectations. This is particularly problematic in environments where labour protections are weak, grievance systems are limited, or managerial authority is already highly centralised. In such settings, AI can inadvertently strengthen asymmetrical power relations by making organisational judgment appear neutral and technical, even when it may reflect incomplete or biased interpretations of worker behaviour (Kondra *et al.*, 2025).

At the same time, AI-driven systems may produce meaningful socioeconomic benefits if deployed responsibly (Sampath *et al.*, 2024). They can help organisations detect burnout earlier, reduce avoidable turnover, improve communication, and support more equitable access to feedback and development opportunities. For employers facing skill shortages, retention challenges, or dispersed workforces, such technologies may contribute to stronger workforce planning and more adaptive management. For employees, well-governed AI systems may improve recognition, reduce invisibility, and provide channels for voicing concerns in settings where direct communication is difficult (Chukwuka and Dibie, 2024). The challenge, therefore, is not whether AI should influence labour management, but how its benefits and burdens are distributed.

These realities make it essential to assess AI adoption not only in terms of efficiency, but also in terms of its consequences for job quality, worker dignity, and inclusion. In African workplaces, the socioeconomic impact of AI-driven labour management will depend on governance choices, organisational intent, and the ability to align digital systems with local labour conditions. Where implementation is transparent, participatory, and supportive, AI may contribute to more responsive and humane workplaces. Where it is extractive, opaque, or punitive, it may intensify disengagement and reproduce the very conditions that drive silent quitting. A critical understanding of these socioeconomic dynamics is therefore essential to any responsible strategy for workplace transformation in the Global South.

## 6. Implementing AI Responsibly: Strategies and Tools

Responsible AI implementation in African workplaces must address not only ethics and compliance, but also the wider transformation of labour relations under digital management systems. This requires organisations to balance efficiency and innovation with worker dignity, inclusion, and context-sensitive governance.

Implementing AI responsibly in Human Resource (HR) practices requires maintaining a critical human oversight component. Organisations should integrate human judgment into AI-driven decision-making processes to enhance perceptions of fairness and effectiveness among employees (Kayusi *et al.*, 2025; Nahar *et al.*, 2025). Human involvement in reviewing and overriding AI-generated decisions when necessary helps mitigate concerns regarding potential biases inherent in automated systems, thereby building trust and acceptance. Human oversight is especially important in high-impact HR matters, such as performance assessment, disciplinary review, promotion decisions, retention decisions, and interventions related to employee well-being. Managers and HR practitioners should review AI-generated insights critically, ask whether the outputs make sense in the local context, and provide space for employees to explain their circumstances when concerns arise. This approach helps prevent overreliance on automated scores or classifications and ensures that context remains central to organisational judgment. In African workplaces, where communication styles, infrastructural conditions, and work arrangements may differ significantly across teams and sectors, this human layer is particularly important for preventing misinterpretation (Chilunjika *et al.*, 2022).

Ethical considerations must be prioritised throughout the development and deployment of AI technologies in HR. Adhering to established ethical frameworks, such as the FEAS principles: Fairness, Explainability, Auditability, and Safety, enables organisations to address challenges like algorithmic bias and transparency (Toreini *et al.*, 2020). Regular audits of AI systems to detect and rectify discriminatory outcomes ensure compliance with ethical standards, which is vital to fostering trust and supporting fair treatment in the workplace. Fairness requires organisations to test whether systems produce unequal outcomes across groups, roles, languages, or demographic categories. Explainability requires that AI outputs can be interpreted and justified in ways that managers and employees can understand. Accountability means that organisations must identify who is responsible for reviewing outputs, correcting errors, and responding to harms when systems fail. Safety involves ensuring that AI tools do not create avoidable risks to employee well-being, dignity, or job security.

Transparency is another cornerstone of responsible AI implementation. Clear communication about how AI tools function and the reasoning behind their decisions is essential for employee understanding and engagement. Providing frequent updates and robust feedback mechanisms helps employees feel informed and involved in the evolving AI-driven HR processes, further enhancing their trust and acceptance of these technologies. Trust is further strengthened when employees see that AI-generated insights lead to meaningful and supportive action. For example, if a pulse survey reveals workload pressure and management responds with realistic workload adjustment or additional support, the system gains legitimacy. If, by contrast, data is collected, but no visible change follows, employees may become more cynical and less willing to engage honestly in future feedback processes. In this sense, responsible AI implementation is not only about technical transparency; it is also about demonstrating that employee voice matters in practice (Malik *et al.*, 2022).

Comprehensive training initiatives play a crucial role in supporting responsible AI use. Targeted training programs should educate employees not only about AI's technical capabilities but also about its ethical dimensions and potential biases. Such training helps demystify AI technologies and empowers staff to engage confidently with AI-based tools, thereby facilitating a culture of transparency and collaboration (Ara and Ahmad, 2025). Organisational readiness is equally critical.

Before implementing AI-driven engagement tools, organisations should assess whether they have the infrastructure, data quality, managerial capacity, and governance processes needed to support effective use. A system introduced into a fragmented or underprepared environment may produce unreliable outputs and create frustration rather than improvement. In many African workplaces, readiness assessments are particularly important because digital maturity often varies significantly between sectors, regions, and institutions (Mamuli *et al.*, 2025). Responsible implementation, therefore, begins with an honest evaluation of organisational capability, not only technological aspiration.

Moreover, fostering active employee involvement in the design, implementation, and evaluation of AI systems encourages a greater sense of ownership and responsibility. When employees perceive that their opinions are recognised and valued, their feelings of fairness and trust towards AI technologies are enhanced (Weber *et al.*, 2022). This participatory approach supports the integration of AI in HR in a manner that aligns with employee expectations and organisational values. Co-design does not require that every employee become a technical expert. Rather, it means creating structured opportunities for workers to shape the design of surveys, reporting channels, communication policies, and governance safeguards. Focus groups, pilot testing, staff consultations, and representative committees can all support this process. In culturally diverse African workplaces, participatory design is especially valuable because it helps ensure that AI systems reflect local realities rather than imported assumptions about how employees communicate or what engagement should look like (Anuyah *et al.*, 2023).

Addressing data privacy and security is imperative for building credible AI systems. Implementing robust data protection measures safeguards sensitive employee information and reassures employees about the safety and integrity of AI-driven decision-making processes. These practices are fundamental for maintaining organisational credibility and employee confidence in AI applications. Responsible practice requires clear data governance rules covering consent, purpose limitation, storage duration, anonymisation, and conditions of access. Employees should know whether their feedback is anonymous, confidential, or identifiable, and be informed whether their data may be used for aggregated reporting, managerial reviews, or future system training. Where possible, organisations should minimise unnecessary data collection and focus on the least intrusive means of achieving legitimate engagement goals.

Responsible AI implementation in African workplaces must be context-sensitive. Technologies developed in other regions often reflect assumptions about language, hierarchy, communication, and work organisation that may not hold in African settings. If these assumptions are not examined, even technically powerful tools may fail to produce meaningful results. Organisations must therefore adapt systems to local realities rather than assume that a single global model of employee engagement can be transferred unchanged across contexts (Anuyah *et al.*, 2023; Bergman *et al.*, 2024). This includes adapting language models to multilingual environments, reviewing the cultural assumptions embedded in survey instruments, and ensuring that feedback mechanisms align with local norms of communication and authority. It also means considering infrastructural factors such as internet reliability, access to devices, and the realities of hybrid or informal work arrangements. In workplaces where direct criticism is culturally sensitive, anonymous channels may be more effective. In settings with strong community values, collective feedback mechanisms may matter more than heavily individualised systems.

Finally, responsible AI implementation requires continuous monitoring and adaptation. Conducting longitudinal research to assess evolving employee perceptions and updating AI tools and governance policies in response to technological advances ensures that AI systems remain effective, fair, and aligned with organisational needs. A strategic and thoughtful approach to integrating AI into HR functions, focusing on augmenting human capabilities such as talent acquisition, performance management, and employee engagement, can optimise benefits while mitigating risks associated with AI adoption (Kayusi *et al.*, 2025; Nahar *et al.*, 2025).

## 7. Conclusion and Recommendations

This paper has explored how AI-driven feedback systems intersect with digital labour and silent quitting in the transformation of African workplaces. It has been shown that silent quitting is not simply an individual problem of low motivation, but a broader response to changing work conditions shaped by burnout, poor leadership, weak recognition, digital work intensification, and uneven organisational support. In African contexts, these pressures are further influenced by infrastructural inequality, limited labour protections, resource constraints, and cultural norms that may discourage open employee voice. The paper argued that AI-driven feedback systems should be understood not only as HR tools, but also as part of the wider automation of labour management. Technologies such as sentiment analysis, pulse surveys, chatbots, engagement dashboards, behavioural analytics, and predictive models are transforming how work is monitored, how employee experience is interpreted, and how managers respond to disengagement. These systems can help organisations identify early signs of withdrawal and improve responsiveness, but they also raise important concerns about privacy, trust, fairness, surveillance, and worker autonomy.

A key insight from this discussion is that the impact of AI in African workplaces depends heavily on context. Digital tools do not operate in a vacuum; their effects are shaped by local labour conditions, organisational culture, leadership capacity, and the level of digital readiness within each workplace. For this reason, the transformation of African workplaces through AI cannot be approached through simple technological transfer. It requires context-sensitive design, strong governance, and a deliberate effort to align innovation with worker dignity and inclusion. Several recommendations follow from this analysis. Organisations should adopt continuous, participatory feedback systems rather than relying solely on periodic engagement reviews. AI-driven tools should be governed by clear rules on consent, data use, accountability, and human oversight. Managers should be trained not only in digital systems, but also in ethical interpretation and empathetic leadership. Policymakers and professional bodies should also develop practical standards for responsible workplace AI that reflect African realities and labour priorities.

Future research should pay closer attention to how AI-mediated labour management is reshaping work across African sectors and occupations. More empirical evidence is needed on whether AI-enabled feedback systems improve engagement and retention over time, and whether they do so fairly across different categories of workers. Overall, transforming African workplaces in the age of AI requires more than introducing new technologies into HR systems. It requires rethinking how digital labour is governed, how employee voice is recognised, and how automation can be used to support trust, fairness, and well-being. When implemented responsibly, AI-driven feedback systems can become valuable tools for addressing silent quitting and building more adaptive, inclusive, and humane workplaces across Africa.